\documentclass{ws-ijns}
\usepackage{graphicx}
\usepackage[english]{babel}
\usepackage[super,sort,compress]{cite}

\begin{document}

\catchline{0}{0}{2005}{}{}

\markboth{Pregowska et al.}{Paper Title}

\title{How far can neural correlations reduce uncertainty? \\
Comparison of Information Transmission Rates for Markov and Bernoulli processes
}

\author{AGNIESZKA PREGOWSKA}

\address{Institute of Fundamental Technological Research, Polish Academy of Sciences, Pawinskiego 5B\\
Warsaw, 02--106, Poland\\
E-mail: aprego@ippt.pan.pl\\
http://www.ippt.pan.pl/staff/aprego}

\author{EHUD KAPLAN}
\address{Icahn School of Medicine at Mount Sinai, One Gustave Levy Place\\
 New York, NY 10029, USA\\
Dept. of Philosophy of Science, Charles University, Prague \\ 
The National Institute of Mental Health, \\
Topolova 748, Klecany, Czechia \\
E-mail: ehud.kaplan@gmail.com\\
http://www.mountsinai.org/profiles/ehud-kaplan}

\author{JANUSZ SZCZEPANSKI}
\address{Institute of Fundamental Technological Research, Polish Academy of Sciences, Pawinskiego 5B\\
Warsaw, 02--106, Poland\\
E-mail: jszczepa@ippt.pan.pl\\
http://bluebox.ippt.pan.pl/~jszczepa/}

\maketitle

\begin{abstract}
The nature of neural codes is central to neuroscience. Do neurons encode information through relatively slow changes in the emission rates of individual spikes (rate code), or by the precise timing of every spike (temporal codes)? Here we compare the loss of information due to correlations for these two possible neural codes.
\par
The essence of Shannon`s definition of information is to combine information with uncertainty: the higher the uncertainty of a given event, the more information is conveyed by that event. Correlations can reduce uncertainty or the amount of information, but by how much? In this paper we address this question by a direct comparison of the information per symbol conveyed by the words coming from a binary Markov source (temporal codes) with the information per symbol coming from the corresponding Bernoulli source (uncorrelated, rate code source). In a previous paper we found that a crucial role in the relation between Information Transmission Rates
$(ITR)$ and Firing Rates is played by a parameter $s$, which is the sum of transitions probabilities from the no-spike-state to the spike-state and \textit{vice versa}. It turned out that also in this case a crucial role is played by the same parameter $s$. We found bounds of the quotient of $ITRs$ for these sources, i.e. this quotient`s minimal and maximal values. Next, making use of the entropy grouping axiom, we determined the loss of information in a Markov source in relation to its corresponding Bernoulli source for a given length of word.
\par
Our results show that in practical situations in the case of correlated signals the loss of information is relatively small, thus temporal codes, which are more energetically efficient, can replace the rate code effectively. These phenomena were confirmed by experiments. 

\end{abstract}


\begin{multicols}{2}
\section{Introduction}
\par
Since the seminal work of Adrian \cite{Adrian_Zotterman_1926} it has been recognized that in the nervous system information is transmitted among spiking neurons by trains of discrete electrical pulses, called action potentials or spikes. A second, non-digital, mode of communication, through gap junctions, is also common (for example,  \cite{Roy2017,Song2016}) but will not be considered here. It has been shown that Firing Rates of spikes change in a consistent manner with inputs. That has given rise to the notion that information is encoded in the neuronal Firing Rate. Recently, in Ref. \cite{Wang_2015} it was presented that in the cerebral cortex various types of neural coding are controlled by the energy field and energy flow. In turn, it is known that Firing Rate is directly related to energy cost \cite{Attwell_Laughlin_2001}, namely, energy consumption increases linearly with spiking frequency \cite{Hasenstaub_2010,Kostal_2013}. On the other hand, many reports show (for example, Ref. \cite{Richmond_1990}) that the total number of spikes varies substantially from trial to trial during the presentation of the same stimulus. This observation has given rise to an alternative hypothesis, which states that additional information is contained in the precise timing of the spikes within the spike train. These two not mutually exclusive views of neural encoding and decoding are broadly categorized as ``rate-based" and ``spike-based" \cite{Brette_2015}.
\par
In Ref. \cite{Dettner_2016} it has been shown that the pairwise temporal spike correlation function within a spike train, and the spike correlation function across repeated presentations of the same stimuli determine the information content in the case of neural codes with finite memory. In Ref. \cite{Mohemmed_2012}, a specific transformation of spike trains into analog signals was applied to explain a mechanism that a spiking neuron is able to learn. It has been argued that associations of arbitrary spike trains in a supervised fashion allow the processing of spatio-temporal information encoded in the precise timing of spikes. 
\par
In our previous papers we compared directly Information Transmission Rates with their corresponding Firing Rates \cite{Pregowska_2015,Pregowska_2016} in the case of binary Information Sources. Our results show that a parameter $s$ (which we called a ``jumping parameter") played a crucial role in the characterization of neural coding. It turned out that depending on this parameter $s$, temporal coding can be more effective than rate coding \cite{Pregowska_2016}. 
\par
In this paper, we compare transmission rates for two types of binary Information Sources: correlated sources and their corresponding independent sources. Making use of the entropy grouping axiom \cite{Ash_1965}, we analyze the relation of information transmitted by the sources described as Markov processes, and by related sources being Bernoulli processes. Our results show that also in this case a crucial role is played by the parameter $s$. We found bounds for the quotients of $ITR$s for these sources, and also their quotients minimal and maximal values. We also determined the loss of Shannon information in Markov sources {\it versus} corresponding Bernoulli sources for a given length of word.
\par
The paper is organized as follows. In Section \ref{entrinfor}, we briefly recall the basic concepts of Information Theory, Bernoulli and Markov processes. In Section \ref{results}, we present the comparison of the $ITR$ of spike trains coming from a Markov source and from the corresponding Bernoulli source. Section Conclusions contains final remarks.
\section{Entropy and Information}\label{entrinfor}
\par
In Shannon`s Theory a communication system is represented by: an input Information Source (stimuli source), a communication channel (neuronal network) and an output Information Source (output signals). In mathematical language sources of information are modelled as stationary discrete stochastic processes. Discrete communication channels are defined by a system of conditional probabilities linking input and output symbols \cite{Shannon_1948,Ash_1965,Cover_Thomas_1991}. In this paper, we study two types of output Information Sources, i.e. sources represented by Markov processes  and by corresponding Bernoulli processes \cite{Feller_1958,Ash_1965}. First, we briefly recall the basic notation \cite{Pregowska_2016}.
\subsection{Entropy}
\par
Let $Z^{L}$ be a set of all words (i.e. blocks) of length $L$, built of symbols (letters) from some finite alphabet $Z$. Each word $z^{L}$ can be treated as a message sent by Information Source $Z$ being a stationary stochastic process. If $P(z^{L})$ denotes the probability the word $z^{L} \in Z^{L}$ occurs, then the information in Shannon sense carried by this word is derived as
\begin{equation}
I(z^{L}):=-\log_{2}{P(z^{L})} \ .\label{infoShanon}
\end{equation}
Thus, the average information of the random variable $Z^{L}$ corresponding to the words of length $L$ is called the Shannon block entropy, and is given by
\begin{equation}
H(Z^{L}):=-\sum\limits_{z^{L}\in Z^{L}}P(z^{L})\log_{2}{P(z^{L})} \ . \label{entrShanon}
\end{equation}
Since the word length $L$ can be arbitrary, the block entropy does not perfectly describe the Information Source \cite{Ash_1965,Cover_Thomas_1991}.
\par
In the special case of a two-letter alphabet $Z=\{0,1\}$ and the length of words $L=1$ we introduce the following notation 
\begin{equation}
H_{2}(p):=H(Z^{1})=-p\log_{2}{p}-(1-p)\log_{2}{(1-p)} \ ,\label{entropy}
\end{equation}
where $P(1)=p$, $P(0)=1-p$ are the associated probabilities. 
\subsection{ Information Transmission and Firing Rates}
\par
The appropriate measure for estimation of transmission efficiency of an Information Source $\textbf{Z}$ is the information transmitted on average by a single symbol, i.e. Information Transmission Rate ($ITR$) \cite{Ash_1965,Cover_Thomas_1991}. 
\par
Let us introduce the notation
\begin{equation}
ITR^{(L)}(\textbf{Z}):= \frac{H(Z^{L})}{L} \label{itr}
\end{equation}
and in the limiting case 
\begin{equation}
ITR(\textbf{Z}):=\lim_{L \to \infty} \frac{H(Z^{L})}{L} \ .\label{itrlimit}
\end{equation}
This limit exists if and only if the stochastic process $\textbf{Z}$ is stationary \cite{Cover_Thomas_1991}.
\par
The most commonly used definition of Firing Rate refers to the temporal average \cite{van_Hemmen_Sejnowski_2006,Crumiller_2013,Gerstner_2014} and is defined as
\begin{equation}
F_{R}=\frac{n_{T}}{T} \ ,\label{fr}
\end{equation}
where $n_{T}$ denotes spike count in a given time window of length $T$ (typically a few seconds). In practice, in order to get sensible averages, some reasonable number of spikes should occur within the time window \cite{Gerstner_2014}. Since the messages are treated as trajectories of locally stationary stochastic process, the Firing Rate as defined by ``Eq. (\ref{fr})"  is specific for a given Information Source provided $T$ is large enough. Thus, $F_{R} \cdot \Delta \tau$ is related to the probability $p$ of spike appearance, where $\Delta \tau$ is the time resolution or bin size. 
\subsection{Information Sources}
\par
An Information Source must produce sequences of symbols, which from a mathematical point of view can be treated as trajectories of a stationary stochastic process $\textbf{Z}= (Z_{i})$, $i=1,2,\dots $ where $Z_{i}$ are random variables \cite{Feller_1958} taking the values from a finite alphabet.
\par
The most commonly used method of digitalization spike trains was proposed in Ref. \cite{MacKay_McCulloch_1952,Bialek_1991,Bialek_1996,DeWesse_1996, Rolls_2004,Levin_2004,Amigo_2004,Pillow_2008,Pregowska_2016}. It is physically justified that spike trains as being observed, are detected with some limited time resolution $\Delta \tau$, so that in each time slice (bin) a spike is either present or absent. If the presence of spike is denoted by ``1" and no spike by ``0", then if we look at some time interval of length $T$, each possible spike train is equivalent to $\frac{T}{\Delta \tau}$ binary sequence which can be treated as trajectory of the stochastic process. 
\par
In Ref. \cite{MacKay_McCulloch_1952,Bialek_1991} it was assumed that random variables which describe the generation of consecutive bits in the sequence representing spike train are independent. This means that these random variables are uncorrelated, i.e. their Pearson Correlation Coefficient $(PCC)$ is equal to 0. Thus, assuming that 1 is generated with probability $p$ (a spike is found in the bin), 0 is generated with probability $1-p$ (a spike is not found), what we have is a Bernoulli process \cite{Feller_1958,Cover_Thomas_1991}.
Clearly, in the case of a Bernoulli process the distribution of $k$ ``ones" between the sequence of bits of length $n$ does not influence the probability of such sequences.  This probability is simply equal, for all such sequences, to $p^{k}(1-p)^{n-k}$ and depends only on the Firing Rate $\frac{k}{n}$. Consequently, since the Shannon information depends only on the probability, all such sequences transmit the same amount of information and we are in the \textit{rate code} regime.
\par
Following the entropy definition ``(\ref{entrShanon})" the Information Transmission Rates ``(\ref{itr})" and ``(\ref{itrlimit})" for Bernoulli process $\textbf{B}$ with the probability of bit 1 equal to $p$ is determined as
\begin{equation}
ITR(\textbf{B}(p))=-p\log_{2}p-(1-p)\log_{2}(1-p) \ .\label{itrb}
\end{equation}
Note that $ITR(\textbf{B}(p))$ is equal to $H_{2}(p)$.
\par
Now, let us assume that the generation of bits of the output signal from an Information Source is described by correlated random variables (in the sense of $PCC$), and this generation is governed by a Markov process $\textbf{M}$.
 In general, a discrete Markov process is defined by a set of conditional probabilities $p_{j|i}$  describing changes from state $i$ to the state $j$, (where $i, j$=0, 1), and by initial distribution probabilities. These changes are called transitions and the probabilities associated with them are called transition probabilities. These probabilities can be put together into a matrix $\textbf{P}$, called the transition matrix, which for the two-states-process is of the form
\begin{equation}
\textbf{P}:= \left[
\begin{array}{ccc}
p_{0|0} & \quad p_{0|1}
\\
&
\\ 
p_{1|0} & \quad p_{1|1}
\end{array}
 \right]=
\left[
\begin{array}{ccc}
1-p_{1|0} & p_{0|1}
\\
&
\\ 
p_{1|0} & 1-p_{0|1}
\end{array}
 \right] \ .   \label{transmatrix} 
\end{equation}
This is a stochastic matrix, i.e. each of its columns sums to 1. Here, we assumed that the process is homogeneous in time. The probability evolution is governed by the Master Equation \cite{van_Kampen_2007}
\begin{equation}
\left[\begin{array}{ccc}
p_{n+1}(0) \\ & \\ p_{n+1}(1)
\end{array} \right]=
\left[\begin{array}{ccc}
1-p_{1|0} & p_{0|1}
\\
&
\\ 
p_{1|0} & 1-p_{0|1}
\end{array}
 \right]\           
\cdot
\left[\begin{array}{ccc}
p_{n}(0) \\ & \\ p_{n}(1)
\end{array} \right]\ , \label{masteq}
\end{equation}   
where $n$ stands for the discrete time, $p_{n}(0)$  and $p_{n}(1)$, are probabilities of finding state ``0" or ``1" at time $n$. In the case of Markov processes the distribution of $k$ ``ones" between the sequence of bits of length $n$ does influence the probability of such sequences. Consequently, since the Shannon information depends only on the probabilities, in general such sequences transmit different amounts of  information. Here the patterns in the sequences of bits do play a role and we are in the \textit{temporal code} regime.
\par
The stationary solution of ``(\ref{masteq})" is given by
\begin{equation}
\left[\begin{array}{ccc}
p_{eq}(0) \\ & \\ p_{eq}(1)
\end{array} \right]=
\left[\begin{array}{ccc}
\frac{p_{0|1}}{(p_{0|1}+p_{1|0})}
\\ & \\
\frac{p_{1|0}}{(p_{0|1}+p_{1|0})}
\end{array} \right]\ . \label{masteqsol}
\end{equation}
The entropy rate ``(\ref{fr})" of the Markov source with transition matrix defined by ``(\ref{transmatrix})" reads \cite{Cover_Thomas_1991} by definition
\begin{equation}
ITR(\textbf{M})= 
\end{equation}
\begin{equation}
P_{eq}(0)[-p_{1|0}\log_{2}{p_{1|0}}-(1-p_{1|0})\log_{2}{(1-p_{1|0})}]+ \nonumber
\end{equation}
\begin{equation}
P_{eq}(1)[-p_{0|1}\log_{2}{p_{0|1}}-(1-p_{0|1})\log_{2}(1-p_{0|1})]\ . \nonumber \label{itrm}
\end{equation}
or, making use of notation ``(\ref{entropy})" it can be written in a compact form
\begin{equation}
ITR(\textbf{M})=p_{eq}(0)H_{2}(p_{1|0})+p_{eq}(1)H_{2}(p_{0|1}) \ . \label{itrm2}
\end{equation}
For the latter, the use of the probability of state ``1" is, in fact, understood as the Firing Rate, and is denoted by $p$,
\begin{equation}
p:=p_{eq}(1)=\frac{p_{1|0}}{(p_{0|1}+p_{1|0})} \ . \label{matrixp}
\end{equation}
For the special case when $p_{0|1}+p_{1|0}=1$, the Markov process $\textbf{M}$ becomes uncorrelated, and reduces to a Bernoulli process with $p=p_{1|0}$.
\par
Under the above notation, we introduced \cite{Pregowska_2016} the ``jumping" parameter $s$ which can be interpreted as the tendency of transition from one state to the other state
\begin{equation}
s:=p_{0|1}+p_{1|0} \ . \label{parametrs}
\end{equation}
Note that $0\leq s \leq 2$.
\par
Using this notation, in the case of the Markov processes $\textbf{M}$, we have
\begin{equation}
p=\frac{p_{1|0}}{s} \ . \label{pe}
\end{equation}
For $0 \leq s \leq 1$, the firing frequency $p$ can take the values from the interval [0, 1], while for $1 \leq s \leq 2$ the values of $p$ are limited to the interval $1-\frac{1}{s} \leq p \leq \frac{1}{s}$. These limits of the range of $p$ follow from ``(\ref{parametrs})", ``(\ref{pe})" and the inequality $s-1 \leq p_{1|0} \leq 1$.
\section{Results}\label{results}
\par
In this Section we compare directly Information Transmission Rates transmitted by spike trains coming from a Markov Information Source as defined by ``(\ref{itrm2})", with $ITR$ of spike trains coming from the corresponding Bernoulli Information Source ``(\ref{itrb})". It is natural to assume that \textit{the Bernoulli process corresponding to a given Markov process} is defined by the stationary probabilities ``(\ref{masteqsol})" of this Markov process. 
\par
\begin{figure*}[t]
\includegraphics[width=3in]{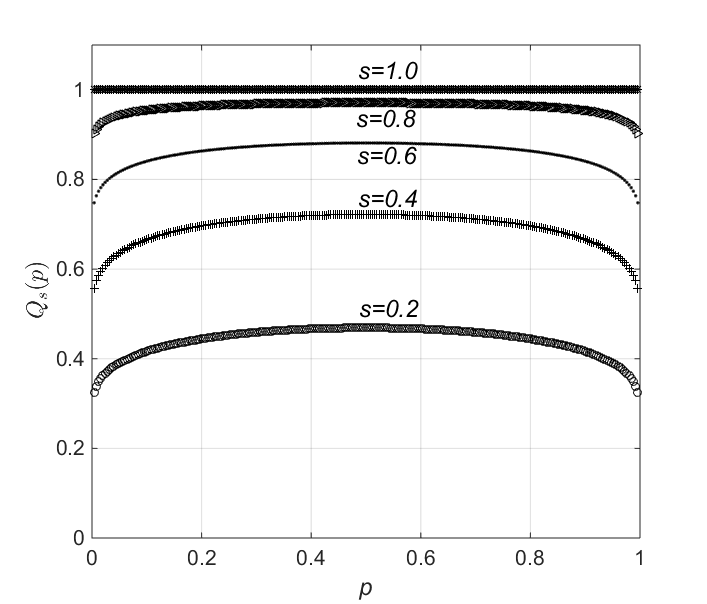}
\includegraphics[width=3in]{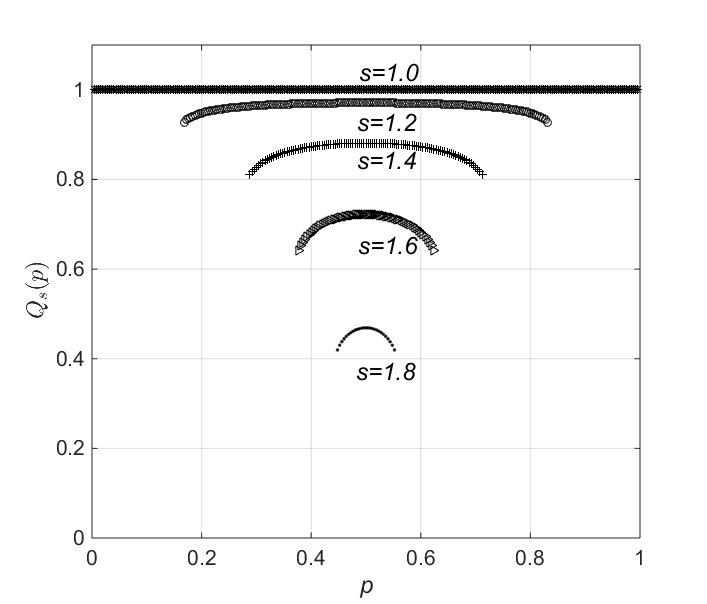} \centering
\caption{The Markov-Bernoulli Information Quotient $Q_{s}$  as a function of Firing Rate  $p$ for a chosen values of the jumping parameter $s$. (A) for parameters $0 \leq s \leq 1$ according to ``(\ref{boun1})" the range of $p$ is [0,1] (B) for parameters $1 \leq s \leq 2$ due to ``(\ref{boun2})" the range of $p$ is $1-\frac{1}{s} \leq p \leq \frac{1}{s}$.}
\label{fig1}
\end{figure*}
\begin{figure*}[t]
\includegraphics[width=3in]{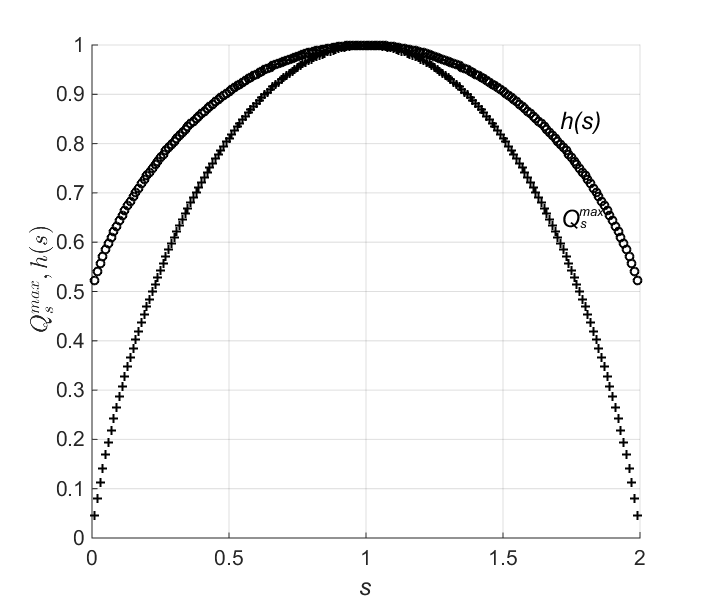} \centering
\caption{The upper bound ``(\ref{h2})" of $Q_{s}$, i.e. $Q_{s}^{max}$ and the limiting factor  $h(s)=\frac{H_{2}\frac{s}{2}+1}{2}$ in ``(33)", ``(34)" as functions of the jumping parameter $s$.}
\label{fig2}
\end{figure*}

Under the notation ``(\ref{itrb})" and ``(\ref{itrm2})", we introduce the following Information Markov-Bernoulli Quotient $Q_{s}$, which is a function of $p$ and depends on parameter $s$ 
\begin{equation}
Q_{s}(p):=\frac{ITR(\textbf{M}_{s}(p))}{ITR(\textbf{B}_{s}(p))} \ . \label{imbq}
\end{equation}
Applying ``(\ref{itrm2})" and ``(\ref{pe})" we have
\begin{equation}
Q_{s}(p):=\frac{(1-p)H_{2}(ps)+pH_{2}((1-p)s)}{H_{2}(p)} \ . \label{imbq2}
\end{equation}
Taking into account the range of $p$, farther on we consider $Q_{s}$ in the following two cases 
\begin{equation}
a) \quad 0 \leq s \leq 1, \quad \mbox{here} \quad 0 \leq p \leq 1 \ , \label{boun1}
\end{equation}
\begin{equation}
b) \quad 1 \leq s \leq 2, \quad \mbox{here} \quad 1-\frac{1}{s} \leq p \leq \frac{1}{s} \ . \label{boun2}
\end{equation}
In Fig.~\ref{fig1} we present, for some arbitrary values of $s$, typical traces of $Q_{s}$ as a function of $p$ both for lower values of the jumping parameter $0 \leq s \leq 1$ (panel A), and for larger values of the jumping parameter $1 \leq s \leq 2$  (panel B). Observe, that $Q_{1}(p)=1$ for each $p$,  due to the fact that for $s$=1, the Markov process reduces to the Bernoulli process. In general, for $s$ close to 1 the amounts of information carried by correlated and corresponding uncorrelated signals are comparable, i.e. the loss of information by correlated signals is relatively small.
Note that $Q_{s}$ for every $s \neq 1$ exhibits one maximum only. One can check the symmetry property i.e.
\begin{equation}
Q_{s}(\frac{1}{2}-r)= Q_{s}(\frac{1}{2}+r)\ . \label{symmetry}
\end{equation}
\begin{equation}
\mbox{for} \quad 0 \leq r \leq \frac{1}{2}, \quad \mbox{in the case of} \quad 0 \leq s \leq 1   \label{limit1}
\end{equation}
and
\begin{equation}
\mbox{for} \quad 0 \leq r \leq \frac{1}{s}-\frac{1}{2}, \quad \mbox{in the case of} \quad 1 \leq s \leq 2 \ . \label{limit2}
\end{equation}
Thus, in both cases, the maximum $Q_{s}^{max}$ over $p$ of the quotient $Q_{s}$ for all values of parameter $s$ is achieved for $p=\frac{1}{2}$ and by ``(\ref{imbq2})" it is equal to
\begin{equation}
Q_{s}^{max}=Q_{s}(\frac{1}{2})=H_{2}(\frac{s}{2})\ . \label{h2}
\end{equation}
In Fig.\ref{fig2}, we show $Q_{s}^{max}$ as a function of s for $0 \leq s \leq 2$.
Here one can observe that the minimal values of the quotient $Q_{s}(p)$ for each $s$ are reached at the endpoints of the intervals ``(\ref{boun1})" and ``(\ref{boun2})".
\par
For $0 \leq s \leq 1$ , making use of ``(\ref{imbq2})", the bounds are as follows 
\begin{equation}
\lim_{p \rightarrow 0} Q_{s}(p)=s \quad \mbox{and} \quad \lim_{p \rightarrow 1} Q_{s}(p)=s\ , \label{lims}
\end{equation}
while for  $1 \leq s \leq 2$ the bounds read
\begin{equation}
\lim_{p \rightarrow 1-\frac{1}{s}} Q_{s}(p)= \frac{\frac{H_{2}(s-1)}{s}}{H_{2}(\frac{s-1}{s})} 
\nonumber
\end{equation}
and by ``(\ref{symmetry})" 
\begin{equation}
\lim_{p \rightarrow \frac{1}{s}} Q_{s}(p)= \frac{\frac{H_{2}(s-1)}{s}}{H_{2}(\frac{s-1}{s})} \ . \label{lims2}
\end{equation}
Observe that for $s \rightarrow 1^{+}$ we have $\frac{\frac{H_{2}(s-1)}{s}}{H_{2}(\frac{s-1}{s})} \rightarrow 1$.
\par
\begin{figure*}[t]
\includegraphics[width=3in]{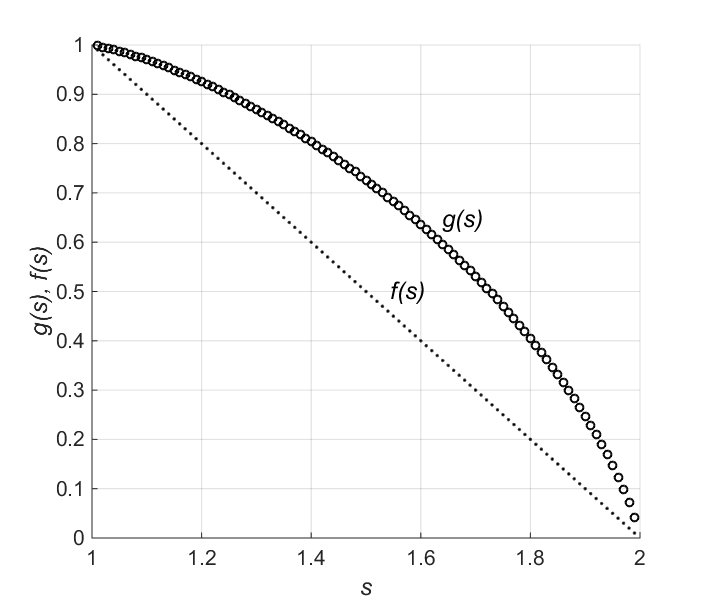} \centering
\caption{The $Q_{s}$' lower bound  $g(s)=\frac{\frac{H_{2}(s-1)}{s}}{H_{2}(\frac{s-1}{s})}$ ``(\ref{lims})" and f(s)=2-s ``(\ref{fs})" as  functions of the  jumping parameter  $1 \leq s \leq 2$.}
\label{fig3}
\end{figure*}

$g(s):=\frac{\frac{H_{2}(s-1)}{s}}{H_{2}(\frac{s-1}{s})}$ as the function of $s$ for $1 \leq s \leq 2$  is shown in Fig.~\ref{fig3}. 
\par
Notice, that for each $s$
\begin{equation}
f(s):=2-s\leq g(s)\ . \label{fs}
\end{equation}
The basic idea of Shannon Information Theory is to combine information with uncertainty. The higher the uncertainty of a given event, the more information is transmitted by such an event. The concept of entropy already addresses this idea. 
\par
To determine how far correlation reduces uncertainty, i.e. in fact the amount of Shannon information, we compare the information per symbol transmitted by the words coming from a binary Markov source with the information per symbol coming from the corresponding Bernoulli source.
\par  
First we consider words of length 2. We make use of the grouping axiom of entropy \cite{Ash_1965}. It is known that the entropy function $H(Z)$ for a discrete random variable $Z$ under assumptions of continuity, monotonicity, uncertainty of joint experiment and grouping axiom, is interpreted as the average uncertainty associated with the events ${Z=z_{i}}$, and is derived as $H(Z)=H(p_{1},\dots,p_{K})= -\Sigma_{j=1}^{K}\log_{2}p_{j}$, where $p_{i}$ is probability of the event ${Z=z_{i}}, i=1,2,\dots,K$.
Let us consider the quotient $Q_{s}$ of the entropy of Markov ``(11)", ``(\ref{itrm2})" and a corresponding Bernoulli process ``(\ref{itrb})". With the above notation and by ``(\ref{entropy})" and ``(\ref{transmatrix})" we have
\begin{equation}
Q_{s}=\frac{p_{eq}(1)H_{2}(p_{0|1})+p_{eq}(0)H_{2}(p_{1|0})}{H_{2}(p_{eq}(1))}= 
\nonumber
\end{equation}
\begin{equation}
\frac{p_{eq}(1)H(p_{0|1},1-p_{0|1})+p_{eq}(0)H(p_{1|0}, 1-p_{1|0})}{H(p_{eq}(1),1-p_{eq}(1))}=
\nonumber
\end{equation}
\begin{equation}
\frac{p_{eq}(1)H(p_{0|1},p_{1|1})+p_{eq}(0)H(p_{1|0}, p_{0|0})}{H(p_{eq}(1),p_{eq}(0))}
\ . \label{qs1}
\end{equation}
Now, we express $Q_{s}$ in the form
\begin{equation}
Q_{s}=\frac{p_{eq}(1)H(\frac{p_{eq}(1)p_{0|1}}{p_{eq}(1)},\frac{p_{eq}(1)p_{1|1}}{p_{eq}(1)})}{H(p_{eq}(1),p_{eq}(0))}+
\nonumber
\end{equation}
\begin{equation}
\frac{p_{eq}(0)H(\frac{p_{eq}(0)p_{1|0}}{p_{eq}(0)},\frac{p_{eq}(0)p_{0|0}}{p_{eq}(0)})}{H(p_{eq}(1),p_{eq}(0))}
 \ . \label{qs2}
\end{equation}
By adding and subtracting $H(p_{eq}(1),p_{eq}(0))$ in the nominator and
making use of the grouping axiom

\begin{equation}
Q_{s}=\frac{p_{eq}(1)H(\frac{p_{eq}(1)p_{0|1}}{p_{eq}(1)},\frac{p_{eq}(1)p_{1|1}}{p_{eq}(1)})}{H(p_{eq}(1),p_{eq}(0))}+
\nonumber
\end{equation}
\begin{equation}
\frac{p_{eq}(0)H(\frac{p_{eq}(0)p_{1|0}}{p_{eq}(0)},\frac{p_{eq}(0)p_{0|0}}{p_{eq}(0)})}{H(p_{eq}(1),p_{eq}(0))}+
\nonumber
\end{equation}
\begin{equation}
\frac{H(p_{eq}(1),p_{eq}(0))-H(p_{eq}(1),p_{eq}(0))}{H(p_{eq}(1),p_{eq}(0))}=
\nonumber
\end{equation}
\begin{equation}
\frac{H(p_{eq}(1)p_{0|1},p_{eq}(1)p_{1|1},p_{eq}(0),p_{1|0},p_{eq}(0)p_{0|0})}{H(p_{eq}(1),p_{eq}(0))}-
\nonumber
\end{equation}
\begin{equation}
\frac{H(p_{eq}(1),p_{eq}(0))}{H(p_{eq}(1),p_{eq}(0))}=
\nonumber
\end{equation}
\begin{equation}
\frac{H(p(1,0),p(1,1),p(0,1),p(0,0))-H(p_{eq}(1),p_{eq}(0))}{H(p_{eq}(1),p_{eq}(0))}
 \ . \label{qs3}
\end{equation}
where $p(i,j)$ denotes the probability of the words $(i,j), i, j =0,1$.
\par
For $0 \leq s \leq 1$ applying ``(\ref{h2})" and ``(\ref{lims})" to ``(\ref{qs3})" we have  
\begin{equation}
s \leq \frac{H(p(1,0),p(1,1),p(0,1),p(0,0))}{H(p_{eq}(1),p_{eq}(0))}-
1 \leq H_{2}(\frac{s}{2})
 \ . \label{res}
\end{equation}
For $1 \leq s \leq 2$ applying ``(\ref{h2})" and ``(\ref{lims2})" to ``(\ref{qs3})" we have
\begin{equation}
\frac{\frac{H_{2}(s-1)}{s}}{H_{2}(\frac{s-1}{s})} \leq 
\nonumber
\end{equation}
\begin{equation}
\frac{H(p(1,0),p(1,1),p(0,1),p(0,0))}{H(p_{eq}(1),p_{eq}(0))}-
1 \leq H_{2}(\frac{s}{2})
 \ . \label{res1}
\end{equation}
and applying ``(\ref{h2})" and ``(\ref{fs})" to ``(\ref{qs3})" we have
\begin{equation}
2-s \leq \frac{H(p(1,0),p(1,1),p(0,1),p(0,0))}{H(p_{eq}(1),p_{eq}(0))}-
1 \leq H_{2}(\frac{s}{2})
 \ . \label{res2}
\end{equation}
Let us consider the $ITR$s for the Markov process against the corresponding Bernoulli process for words of length 2. Making use of ``(\ref{res})" and ``(\ref{res2})", and using the notation ``(\ref{itr})", we obtain the following relation between these Information Sources:
\begin{equation}
\frac{s+1}{2}ITR(\textbf{B}_{s}(p)) \leq ITR^{(2)}(\textbf{M}_{s}(p)) \leq
\nonumber
\end{equation}
\begin{equation}
 \frac{H_{2}(\frac{s}{2})+1}{2}ITR(\textbf{B}_{s}(p)) 
\nonumber
\end{equation}
\begin{equation}
\quad \mbox{for} \quad 0 \leq s \leq 1
 \ , \label{res3}
\end{equation}
and
\begin{equation}
\frac{3-s}{2}ITR(\textbf{B}_{s}(p)) \leq ITR^{(2)}(\textbf{M}_{s}(p)) \leq 
\nonumber
\end{equation}
\begin{equation}
\frac{H_{2}(\frac{s}{2})+1}{2}ITR(\textbf{B}_{s}(p)) 
\nonumber
\end{equation}
\begin{equation}
\quad \mbox{for} \quad 1 \leq s \leq 2
 \ , \label{res4}
\end{equation}
where $ITR^{(2)}(\textbf{M}_{s}(p))$ denotes the Information Transmission Rate ``(\ref{itr})" of the Markov process for words of length 2, and $ITR(\textbf{B}_{s}(p))=\frac{1}{2}ITR^{(2)}(\textbf{B}_{s}(p))$ is Information Transmission Rate of the corresponding Bernoulli process. Note that the correlation can reduce the $ITR$ by as much as a half, i.e
\begin{equation}
\frac{1}{2}ITR(\textbf{B}_{s}(p)) \leq ITR^{(2)}(\textbf{M}_{s}(p)) \leq ITR(\textbf{B}_{s}(p)) 
\nonumber
\end{equation}
\begin{equation}
\quad \mbox{for} \quad 0 \leq s \leq 1
 \ , \label{res5}
\end{equation}
\begin{equation}
\frac{1}{2}ITR(\textbf{B}_{s}(p)) \leq ITR^{(2)}(\textbf{M}_{s}(p)) \leq ITR(\textbf{B}_{s}(p)) 
\nonumber
\end{equation}
\begin{equation}
\quad \mbox{for} \quad 1 \leq s \leq 2
 \ . \label{res6}
\end{equation}
Similar considerations for words of length $n\quad(n \geq 2)$ led to the more general formulas
\begin{equation}
[s(1-\frac{1}{n})+\frac{1}{n}]ITR(\textbf{B}_{s}(p)) \leq ITR^{(n)}(\textbf{M}_{s}(p)) \leq 
\nonumber
\end{equation}
\begin{equation}
[H_{2}(\frac{s}{2})(1-\frac{1}{n})+\frac{1}{n}]ITR(\textbf{B}_{s}(p)) 
\nonumber
\end{equation}
\begin{equation}
\quad \mbox{for} \quad 0 \leq s \leq 1
 \ , \label{res7}
\end{equation}
\begin{equation}
[(2-s)(1-\frac{1}{n})+\frac{1}{n}]ITR(\textbf{B}_{s}(p)) \leq ITR^{(n)}(\textbf{M}_{s}(p)) \leq 
\nonumber
\end{equation}
\begin{equation}
[H_{2}(\frac{s}{2})(1-\frac{1}{n})+\frac{1}{n}]ITR(\textbf{B}_{s}(p)) 
\nonumber
\end{equation}
\begin{equation}
\quad \mbox{for} \quad 1 \leq s \leq 2
 \ . \label{res8}
\end{equation}
Note, that from ``(\ref{res7})" and ``(\ref{res8})" the following upper and lower bounds follow
\begin{equation}
sITR(\textbf{B}_{s}(p)) \leq 
ITR^{(n)}(\textbf{M}_{s}(p)) \leq 
\nonumber
\end{equation}
\begin{equation}
\frac{H_{2}(\frac{s}{2})+1}{2}ITR(\textbf{B}_{s}(p)) 
\nonumber
\end{equation}
\begin{equation}
\quad \mbox{for} \quad 0 \leq s \leq 1
 \ , \label{res9}
\end{equation}
\begin{equation}
(2-s)ITR(\textbf{B}_{s}(p)) \leq 
ITR^{(n)}(\textbf{M}_{s}(p)) \leq 
\nonumber
\end{equation}
\begin{equation}
\frac{H_{2}(\frac{s}{2})+1}{2}ITR(\textbf{B}_{s}(p)) 
\nonumber
\end{equation}
\begin{equation}
\quad \mbox{for} \quad 1 \leq s \leq 2
 \ , \label{res10}
\end{equation}
where in Ref. \cite{Cover_Thomas_1991} $ITR(\textbf{B}_{s}(p))=H_{2}(p)$ and $n \geq 2$. Note, that the bounds $s$ and $2-s$ can be interpreted as $1-\mbox{det}\bf{P}$ and $\mbox{tr}\bf{P}$, respectively.
\par 
These results show that for Markov processes for any length of word, the reduction of information due to correlations is limited by the factor $s$ or $2-s$. This finding supports the hypothesis that under certain conditions neurons can use temporal codes which are more energetically efficient compared to the more reliable rate code. It is interesting that the factors (bounds) in the above inequalities depend only on the jumping parameter $s$. This parameter is simply the sum of the conditional probabilities of transition from state to state. On the other hand, experiments show \cite{Strong_1998} that spiking frequency is in practice limited typically by 40 spikes within a time period of a few seconds and time resolution of the spikes being detected is typically in the range of 3 ms. Thus, it is justified to assume that, after digitalization, the transition probability from the state in which there is a spike, to the state where there is no-spike is large (i.e. close to 1), while the transition probability from the state of no-spike to the state where there is a spike is small (i.e. close to 0), and consequently the values of $s$ are around 1. However, our results show that for $s$ close to 1, the amounts of information carried by correlated (like temporal codes) and corresponding uncorrelated (like rate code) the signals are comparable. This suggests that when a neuronal system decides to use a temporal code, some trade-off between energetic cost and transmission reliability must be taken into account. Experiments confirm that such situations can occur in the primary auditory cortex \cite{deCharms_Merzenich_1996,Trussell_1997,Oertel_1999, Wang_2008}, the visual cortex, \cite{Victor_Purpura_1996} and also in the olfactory \cite{Laurent_1997} and the gustatory \cite{DiLorenzo_2009} information processing systems.  
\section{Conclusions}
Spiking neurons communicate with each other by means of small electric currents, transferring information via sequences of action potentials called spike trains, which can be viewed as a string of binary signals \cite{MacKay_McCulloch_1952}. It is still an open question whether the information contained in these binary signals is conveyed by the firing frequency, or by the precise timing of the spikes. The nature of the code used by spike trains is closely related to whether the digitalized representation of messages is governed by {\em uncorrelated} stochastic processes (Bernoulli processes) or by {\em correlated} ones, such as some Markov processes. 
\par
We point out that the correlations we have considered in this paper refer {\em only} to correlations within a given spike train, and are thus distinct from correlations among spike trains emitted by several different neurons, a topic that has received a great deal of experimental and theoretical attention (for example, \cite{averbeck2006,cohen2011,Eyherabide2013}).
\par
In this paper we have shown, that when information conveyed by spike trains coming from such different sources is compared, a crucial role is played by the same jumping parameter $s$, as we found in Ref. \cite{Pregowska_2016}. We have found that the correlation-related loss of information for signals governed by Markov processes, when compared with the corresponding uncorrelated processes, is determined only by this parameter $s$. Experiments confirm that taking into account the frequency of neuronal signals and spike detection resolution this parameter oscillates around 1. Our results show that for $s$ close to 1, the amounts of information transmitted by correlated and corresponding uncorrelated signals are comparable. Thus temporal codes, which are more energetically efficient, can be used instead of rate codes. This was observed in a number of \textit{in vivo} recordings of neuronal activity \cite{Baker_2013} and in the studies mentioned in that reference.

\nonumsection{Acknowledgments} \noindent This paper has been partly conceived through participation in the EMBO Installation Grant of dr Michal Komorowski. Thus, we thank MK for valuable discussions.

\end{multicols}

\end{document}